\documentstyle[sprocl]{article}

\input{psfig}

\def\Journal#1#2#3#4{{#1} {\bf #2}, #3 (#4)}

\def\ANN{\em Ann. Phys.}
\def\ANP{\em Adv. Nucl. Phys.}
\def\JPG{{\em J. Phys.} G}

\def\PLB{{\em Phys. Lett.} B}
\def\PRL{\em Phys. Rev. Lett.}
\def\PRC{{\em Phys. Rev.} C}
\def\PRD{{\em Phys. Rev.} D}
\def\PRP{\em Phys. Rep.}
\def\ZPA{{\em Z. Phys.} A}
\def\ZPC{{\em Z. Phys.} C}

\begin{document}

\hspace*{7cm} JLAB-THY-99-26

\hspace*{7cm} ADP-99-34/T371	\\ \\

\title{FLAVOR DECOMPOSITION OF THE NUCLEON
\footnote{Extended version of talk presented at the Workshop on
	Exclusive \& Semi-Exclusive Processes at High Momentum Transfer,
	Jefferson Lab, June 1999.}}

\author{W. MELNITCHOUK}
\address{Jefferson Lab, 12000 Jefferson Avenue,
	Newport News, VA 23606, \\
	and Special Research Centre for the
	Subatomic Structure of Matter,  \\ 
	University of Adelaide, Adelaide 5005, Australia}

\maketitle

\abstracts{I review some recent developments in the study of
quark flavor distributions in the nucleon, including
(i) valence quark distributions and the quark--hadron duality
prediction for the $x \to 1$ $d/u$ ratio
(ii) sea quark asymmetries and electromagnetic form factors
(iii) strange quarks in the nucleon.}

%%%%%%%%%%%%%%%%%%%%%%%%%%%%%%%%%%%%%%%%%%%%%%%%%%%%%%%%%%%%%%%%%%%%%%%%%
\section{Introduction}

While the problem of how the proton's {\em spin} is distributed among
its constituents continues to captivate the attention of a large
segment of the hadron physics community~\cite{LR} and stimulate
development of new approaches to the problem~\cite{NFPD}, the question
of how the proton's momentum is distributed among its various
{\em flavors} is far from being satisfactorily answered.
Recent experiments and refined data analyses have indeed forced us
to go far beyond the naive view of a nucleon as three non-relativistic
valence quarks in a sea of perturbatively generated $q\bar q$ pairs
and gluons~\cite{SCIENCE}.

A classic example of this is the asymmetry of the light quark sea,
dramatically confirmed in the recent Drell-Yan experiment at
Fermilab~\cite{E866}, whose interpretation defies any perturbative
understanding~\cite{DYN}.
In a self-consistent representation of the nucleon, the dynamics
responsible for any non-perturbative effects visible in quark
distributions should also leave traces in other flavor-sensitive
observables such as electromagnetic form factors.
One can quantitatively study the connection between the quark
distributions and form factors in the context of non-forward parton
distributions, which is one of the offshoots of the recent proton spin
decomposition studies~\cite{NFPD}.
As discussed in Section 3, an important test of any realistic model
of nucleon structure is that it be able to account not only for the
asymmetries in sea quark distributions such as the $\bar d/\bar u$
ratio, but also observables such as the electric form factor of neutron,
which is particularly sensitive to the spin-flavor dynamics of quarks.

Less firmly established, but quite likely to exist nonetheless,
are asymmetries between quark and antiquark distributions for
heavier flavors, such as $s$ and $\bar s$, and even $c$ and $\bar c$,
which are discussed in Section 4.
These are closely connected with the strangeness form factors of the
nucleon which are currently receiving much attention from theory
and experiment alike.

At the same time as the proton continues to reveal a rich substructure
of its sea, some important details of quark distributions in the valence
region still remain elusive.
Most conspicuous of these is the valence $d$ quark distribution,
or the $d/u$ ratio, whose $x \rightarrow 1$ limit remains
controversial~\cite{NP,MRST}.
A number of proposals~\cite{OTHER} have been made recently for determining
the large-$x$ behavior of $d/u$ in semi-inclusive deep-inelastic
scattering and other high-energy processes.
In Section 2 I recall an old prediction for the $d/u$ ratio at large $x$
based on empirical observations of quark--hadron duality first made
nearly 3 decades ago.

%%%%%%%%%%%%%%%%%%%%%%%%%%%%%%%%%%%%%%%%%%%%%%%%%%%%%%%%%%%%%%%%%%%%%%%%%
\section{Valence Quarks and Quark--Hadron Duality}

The valence $d/u$ ratio contains important information about the
spin-flavor structure of the proton~\cite{ISGUR_V}, and its asymptotic
$x \rightarrow 1$ behavior reflects the mechanism(s) responsible for the
breaking of SU(2)$_{\rm spin} \times$ SU(2)$_{\rm flavor}$ symmetry.
There are a number of predictions for this ratio, ranging from 1/2
in the non-relativistic SU(6) quark model, to 0 in broken SU(6) with
scalar-isoscalar spectator quark dominance~\cite{BRSU6}, to 1/5 in
perturbative QCD~\cite{FJ}.
Indeed, this is one of the very few predictions for the $x$-dependence
of parton distributions which can be drawn from perturbative QCD, 
and its verification or failure would be an important indicator of
the appropriate kinematics at which QCD can be treated perturbatively.

The biggest obstacle to an unambiguous determination of $d/u$ at
large $x$ is the fact that this ratio is extracted from the ratio
of inclusive neutron and proton structure functions, with the former
never measured directly but rather inferred from proton and deuteron
cross sections.
The deuteron cross sections, however, must be corrected for nuclear
effects in the structure function, which can become quite
significant~\cite{FS,SMEAR} at large $x$.
In particular, whether one corrects for Fermi motion only, or in addition
for binding and nucleon off-shell effects, the extracted neutron structure
function for $x > 0.7$ can differ dramatically~\cite{NP}.
The original observation~\cite{NP} that the $d/u$ ratio, when corrected
for nuclear effects in deuterium, is larger than the value assumed in 
global data parameterizations was confirmed recently in a subsequent
reanalysis~\cite{BODEK} based on the assumption that the nuclear
corrections scale with nuclear density~\cite{FS}.
While this may be a reasonable assumption for heavy nuclei such as
$^{56}$Fe or $^{40}$Ca, it is rather speculative when applied to
light nuclei such as the deuteron.
Nevertheless, the conclusions of both analyses do suggest that the
neutron structure function may be significantly underestimated through
the neglect of nuclear effects.

\begin{figure}[ht]
\hspace*{2cm}
\psfig{figure=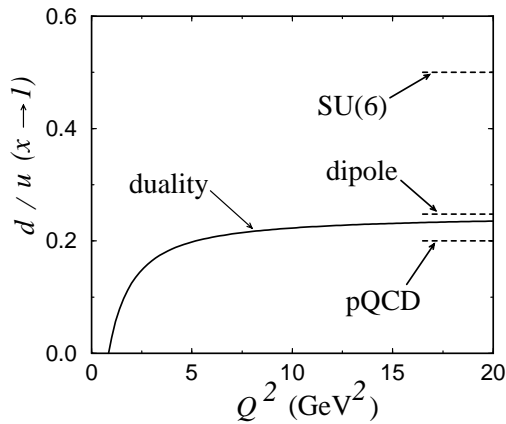,height=5.5cm}
\caption{Quark--hadron duality prediction for the $x \rightarrow 1$
	behavior of the $d/u$ ratio.}
\end{figure}

While a number of suggestions have been made how to avoid the nuclear
contamination problem~\cite{OTHER}, one of the more direct ways is to
measure relative yields of $\pi^+$ and $\pi^-$ mesons in semi-inclusive
deep-inelastic scattering from protons~\cite{SEMIPI}.
At large $z$ ($z$ being the energy of the pion relative to the photon)
the $u$ quark fragments primarily into a $\pi^+$, while a $d$
fragments into a $\pi^-$, so that at large $x$ the ratio
$R^\pi = \sigma(\pi^-) / \sigma(\pi^+)$ is given by the ratio
$d/u$ weighted by $q \rightarrow \pi$ fragmentation functions.
Indeed, in the limit $z \rightarrow 1$, where the $u \rightarrow \pi^+$
fragmentation function dominates~\cite{SEMIPI}, the ratio
$R^{\pi} \rightarrow d/4u$.

A direct semi-inclusive measurement of fast pion production will require
relatively large $Q^2$ and $W^2$, which may not be feasible until a
facility such as an upgraded 12 GeV electron beam at Jefferson Lab  
becomes available.
In the meantime, one must look for clues from other sources for
information about $d/u$ at large $x$, and one of the more obscure
ones is the phenomenon of quark--hadron duality in inelastic structure
functions.

As observed originally by Bloom and Gilman~\cite{BG}, when averaged
over some interval of $\omega' = (2 M \nu + M^2)/Q^2$, or more
precisely~\cite{RUJ} the Nachtmann scaling variable
$\xi = 2x / (1 + \sqrt{1 + 4M^2x^2/Q^2})$, the inclusive structure    
function in the resonance region at low $W$ is approximately equal
to the scaling structure function at much larger $Q^2$.
This was later reinterpreted by de R\'ujula, Georgi and  
Politzer~\cite{RUJ} in terms of an operator product expansion of  
the Nachtmann moments of $F_2$, in which the equality of the low   
moments was understood to arise from the relatively small size of  
higher twist ($1/Q^2$ suppressed) contributions compared with the
leading twist.

Recent experiments at Jefferson Lab~\cite{JLAB} confirm that this
observation is reasonably accurate for each of the low-lying resonances,
including the extreme case of elastic scattering.
If one takes the latter seriously, then the first moment of the elastic
structure function of the proton is given by the electromagnetic form
factors~\cite{RUJ,RIC}:
\begin{eqnarray}
\label{elastic}
\int_{\xi_{th}}^1 d\xi\ F_2^{\rm el}(\xi,Q^2)
&=& {\xi_0^2 \over 2 - \xi_0} {\cal G}(Q^2),
\end{eqnarray}
where $\xi_{th}$ is the value of $\xi$ at the pion threshold, and
\begin{eqnarray}
{\cal G}(Q^2)
&=& { 1 \over 1 + \tau }
\left( G_E^2(Q^2) + \tau G_M^2(Q^2) \right),
\end{eqnarray}
with $\tau = Q^2/4M^2$ and $\xi_0 = 2 / (1 + \sqrt{1 + 1/\tau})$ is
the value of $\xi$ at $x=1$.
Differentiating both sides of Eq.(\ref{elastic}) with respect to $Q^2$
gives~\cite{BG}:
\begin{eqnarray}
\label{rat}
\left. { F_2^n(x,Q^2) \over F_2^p(x,Q^2) }
\right|_{x \rightarrow 1}
&=& { d{\cal G}^n(Q^2)/dQ^2 \over d{\cal G}^p(Q^2)/dQ^2 },
\end{eqnarray}
where
\begin{eqnarray}
{ d{\cal G} \over dQ^2 }
&=& { G_M^2 - G_E^2 \over 4 M^2 (1+\tau)^2 }
 + { 1 \over (1+\tau) }
   \left( { dG_E^2 \over dQ^2 } + \tau { dG_M^2 \over dQ^2 }
   \right).
\end{eqnarray}  
Using empirical values for the form factors and inverting
$F_2^n/F_2^p$ gives the local duality prediction for the $d/u$
ratio in the $x \rightarrow 1$ limit in Fig.1.
If one further assumes that both the proton and neutron magnetic
form factors have the same dipole form at large $Q^2$, then
$d/u \rightarrow (\mu_p^2 - 4 \mu_n^2)/(\mu_n^2 - 4 \mu_p^2)
     \approx 0.25$ in the limit $Q^2 \rightarrow \infty$.

It's interesting to observe that above $Q^2 \approx 5$ GeV$^2$ the
duality prediction is similar numerically to the expectation from
perturbative QCD.
This is consistent with the operator product expansion interpretation
of de R\'ujula et al.~\cite{RUJ} in which duality should be a better   
approximation with increasing $Q^2$.
Whether this is a coincidence or an indicator of common underlying
physics remains to be settled by future experiments.

%%%%%%%%%%%%%%%%%%%%%%%%%%%%%%%%%%%%%%%%%%%%%%%%%%%%%%%%%%%%%%%%%%%%%%%%%
\section{Role of the Light Quark Sea}

The $\bar d/\bar u$ ratio is an important testing ground for
our ideas about the long-range structure of the nucleon, and the
origin of the non-perturbative spin-flavor interaction.
The latest round of discussion about the proton's $\bar d$ and $\bar u$
distributions has been spurred on by the recent measurement by the E866
Collaboration~\cite{E866} at Fermilab of the $x$-dependence of the $pd$
to $pp$ cross section ratio for Drell-Yan production, which is sensitive
to the $\bar d/\bar u$ ratio at small $x$.

\begin{figure}[ht]
\hspace*{1.5cm}
\psfig{figure=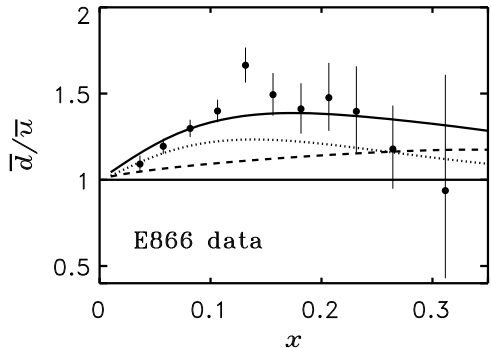,height=5.5cm}
\caption{Flavor asymmetry of the light antiquark sea, including pion
	cloud (dashed) and Pauli blocking effects (dotted), and the
	total (solid)~\protect\cite{DYN}.}
\end{figure}

As pointed out originally by Field and Feynman~\cite{FF}, because the
valence quark flavors are unequally represented in the proton, the Pauli
exclusion principle implies that $\bar u u$ pair creation is suppressed
in the proton relative to $\bar d d$.
Later, Thomas~\cite{AWT83} observed that an excess of $\bar d$ quarks
in the proton also arises naturally from the chiral structure of QCD,
in the form of a pion cloud.
In simple terms, if part of the proton's wave function has overlap with
a virtual $\pi^+ n$ state, a deep-inelastic probe scattering from the
virtual $\pi^+$, which contains a valence $\bar d$ quark, will
automatically lead to $\bar d > \bar u$ in the proton.

Whatever the ultimate origin of the asymmetry, it is likely to involve
some non-perturbative spin-flavors interaction between quarks.
In order to identify the different possible origins of the asymmetry,
consider a model of the nucleon in which the nucleon core consists of
valence quarks, possibly interacting via exchange of gluons, with sea
quark effects introduced through the coupling of the core to $q \bar q$
states with light meson quantum numbers (many variants of such a model
exist --- see for example Refs.\cite{CBM,GI}).
In practice, because of their pseudo-Goldstone nature and their
anomalously light mass, the pseudoscalar pions (and for strange
observables, kaons) are the most important $q \bar q$ states.
The effects of the Pauli exclusion principle on the $\bar d/\bar u$
ratio can then be associated with antisymmetrization of $q \bar q$
pairs created inside the core~\cite{FMST}.

Figure 2 shows the $\bar d/\bar u$ ratio in the proton including flavor
symmetry breaking effects generated by a pion cloud, with both $N$ and
$\Delta$ recoil states (for soft hadronic form factors the contributions
from heavier mesons and baryons are small~\cite{MTVEC,REV}), and a
contribution due to the Pauli blocking effect~\cite{DYN}.
The pion cloud parameters, namely the hadronic $\pi NN$ and $\pi N\Delta$
vertex form factors, are taken from the measured values of the axial
elastic $N$ and $N \Delta$ transition form factors~\cite{AXIAL}, and give
an overall pion probability in the proton of $\approx 10-15\%$.

\begin{figure}[h]
\hspace*{2cm}
\psfig{figure=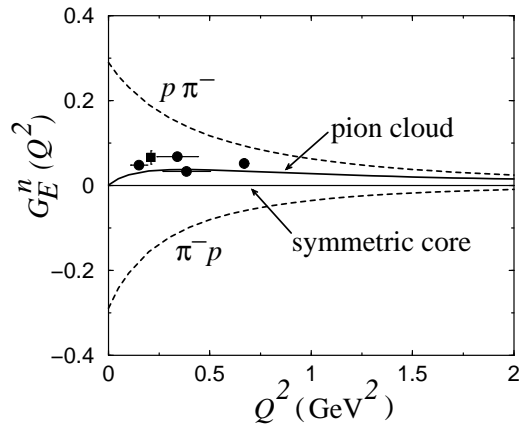,height=5.5cm}
\caption{Neutron electric form factor in the pion cloud model.}
\end{figure}

If a pseudoscalar cloud of $q \bar q$ states plays an important role
in the $\bar d/\bar u$ asymmetry, its effects should also be visible
in other flavor-sensitive observables, such as electromagnetic form
factors.
An excellent example is the electric form factor of the neutron,
a non-zero value for which can arise from a pion cloud,
$n \rightarrow p \pi^-$.
Although in practice other effects~\cite{IKS,ISGUR_G} such as 
spin-dependent interactions due to one gluon exchange between core
quarks may also contribute to $G_E^n$, it is nevertheless important 
to test the consistency of the above model by evaluating its
consequences for all observables that may carry its signature.

To illustrate the sole effect of the pion cloud, all residual
interactions between quarks in the core are switched off,
so that $G_E^n$ has only two contributions: one in which the photon
couples to the virtual $\pi^-$ (labeled ``$\pi^- p$'' in Fig.3)
and one where the photon couples to the recoil proton (``$p \pi^-$''
in Fig.3).
Both contributions are large in magnitude but opposite in sign,
so that the combined effects cancel to give a small positive
$G_E^n$, consistent with the data.

One should stress that {\em the same} pion cloud parameters are   
used in the calculation of $G_E^n$ as for the $\bar d/\bar u$ 
asymmetry in Fig.4.
Note, however, that the Pauli blocking effect plays no role in
form factors, since any suppression of $\bar u$ relative to $\bar d$
here would be accompanied by an equal and opposite suppression of
$u_{\rm sea}$ relative to $d_{\rm sea}$, and form factors measure
charge conjugation odd (i.e. valence) combinations of flavors.
The fact that the model prediction slightly underestimates the strength
of the observed $G_E^n$ suggests that other mechanisms, such as one
gluon exchange between quarks in the core~\cite{IKS}, could be
responsible for some of the difference.

\begin{figure}[h]
\hspace*{2cm}
\psfig{figure=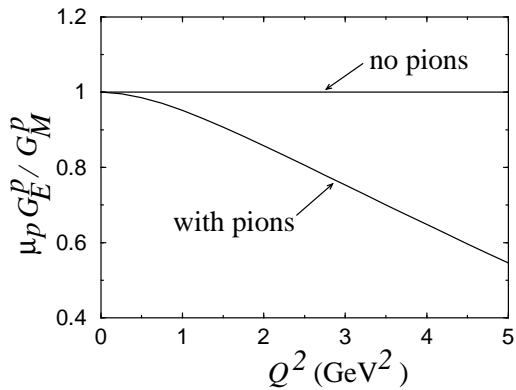,height=5cm}
\caption{Electric to magnetic proton form factor ratio,
	with and without a pion cloud.}
\end{figure}

Another quantity sensitive to details of the flavor distributions
in the proton is the electric to magnetic form factor ratio,
which was recently measured at Jefferson Lab~\cite{GEM}.
Perturbative QCD predicts that asymptotically this ratio should be
$Q^2$-independent~\cite{BROD}, so that any deviation from a constant
ratio would be due to the influence of non-perturbative dynamics.
Again assuming a symmetric core which leaves $\mu_p G_E^p = G_M^p$
for all $Q^2$, the pion cloud contribution to the Pauli form factor
is suppressed in $G_E^p$ relative to that in $G_M^p$, resulting in
the softening of the $G_E^p/G_M^p$ ratio.
Compared with a dipole parameterization, the pion corrections leave
$G_M^p$ relatively unaffected, but make $G_E^p$ softer, leading to
the ratio in Fig.4.

More quantitative comparisons with data would require one to consider
more sophisticated models of the nucleon, incorporating explicitly
the dynamics of core quarks as well as $q \bar q$ pairs.
These simple examples, however, serve to illustrate the point that the
structure and interactions of light quarks and antiquarks in both
electromagnetic form factors and high-energy scattering are far
richer than could ever be inferred from perturbative QCD.

%%%%%%%%%%%%%%%%%%%%%%%%%%%%%%%%%%%%%%%%%%%%%%%%%%%%%%%%%%%%%%%%%%%%%%%%%
\section{Strange Quarks in the Nucleon}

A complication in studying the light quark sea is the fact that 
non-perturbative features associated with $u$ and $d$ quarks are
intrinsically correlated with the valence core of the proton,
so that effects of $q \bar q$ pairs can be difficult to distinguish
from those of antisymmetrization.
The strange sector, on the other hand, where antisymmetrization
between sea and valence quarks plays no role, is therefore more
likely to provide direct information about the non-perturbative
origin of the nucleon sea~\cite{JI}.

\begin{figure}[ht]
\hspace*{1.5cm}
\psfig{figure=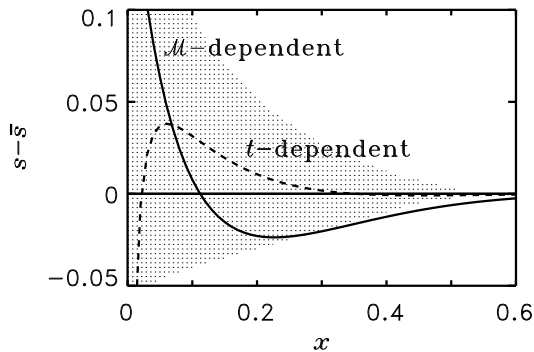,height=5.5cm}
\caption{Strange quark asymmetry in the proton arising from meson
	clouds for two different $KNY$ form factors.
	The shaded region indicates current experimental limits
	from the CCFR Collaboration~\protect\cite{CCFR}.}
\end{figure}

Two related observables which would indicate the presence of
non-perturbative strange quarks are the $s - \bar s$ quark
distribution asymmetry, and strange electromagnetic form factors.
Limits on the former have been obtained from charm production cross
sections in $\nu$ and $\bar\nu$ deep-inelastic scattering, which probes
the $s$ and $\bar s$ distributions in the nucleon, 
respectively~\cite{CCFR}.
The resulting difference $s-\bar s$, indicated in Fig.5 by the shaded
area, is consistent with zero, but also consistent with a small amount
of non-perturbative strangeness, which would be generated from a kaon
cloud around the nucleon~\cite{MM}, as in chiral quark models or SU(3)
chiral perturbation theory.
Indeed, the simplest models of intrinsic strangeness in the nucleon assume
that the strangeness is carried by its $KY$ ($Y=\Lambda, \Sigma, \cdots$)
components, so that the $s$ and $\bar s$ quarks have quite different
origins~\cite{GI,ST}.

The kaon cloud contribution to the asymmetry is shown by the solid
curve in Fig.5, for a kaon probability of $\approx 3\%$.
Because the $\bar s$ distribution in a kaon is much harder than the
$s$ distribution in a hyperon, the resulting $s-\bar s$ difference
will be negative at large $x$, despite the kaon distribution in the
nucleon being slightly softer than the hyperon distribution (on the
light cone).
However, contrary to recent claims in the literature, a kaon cloud
does not unambiguously predict the sign of the $s-\bar s$ difference
as a function of $x$, which turns out to be very sensitive to the
dynamics of the $KNY$ vertex~\cite{MM}.
To demonstrate this the asymmetry is calculated for two different $KNY$
form factors, one which depends on the invariant-mass ${\cal M}$ of the
$KY$ state~\cite{REV,MM} (solid), and one which depends on the exchanged
four-momentum $t$ (dashed).
For the latter, the $K$ distribution in the nucleon is somewhat softer
than the $\Lambda$, which in this case does overcompensate for the harder
$\bar s$ distribution in $K$~\cite{MM}.
Better precision neutrino data would therefore be extremely helpful
in determining just how asymmetric strangeness in the nucleon is.

\begin{figure}[h]
\hspace*{1.5cm}
\psfig{figure=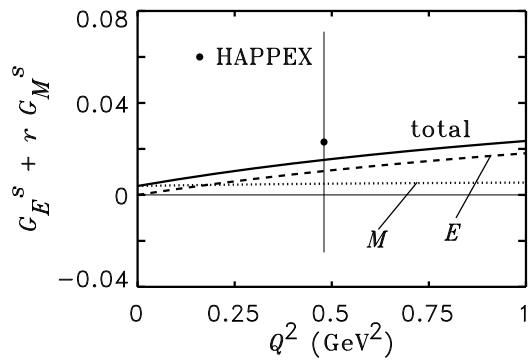,height=5.5cm}
\caption{Strange electromagnetic form factors of the proton
	compared with a kaon cloud prediction.
	For the HAPPEX data~\protect\cite{HAPPEX}, $r\approx 0.4$.}
\end{figure}

The other way to determine the size of the non-perturbative strange
asymmetry in the nucleon is to measure the strange contributions to
the elastic electromagnetic form factors, as has recently been done
at MIT-Bates~\cite{SAMPLE} and Jefferson Lab~\cite{HAPPEX}.
The latest experimental information on the strange electric and magnetic
form factors is shown in Fig.6, together with the kaon cloud prediction
using the same parameters for the $KNY$ vertex as in the $s-\bar s$
asymmetry in Fig.5.
The result is a small and slightly positive~\cite{MM} value for
the $G_E^s$ and $G_M^s$ combination measured in the HAPPEX
experiment\footnote{The result of a new measurement of the $G_M^s$
form factor just announced by the SAMPLE Collaboration~\cite{SAMPLE_NEW}
gives a value $+0.61 \pm 0.17 \pm 0.21$ at $Q^2=0.1$ GeV$^2$, providing
the strongest evidence yet for a non-zero strange asymmetry in the 
proton.}.
Although in good agreement with the available data, clearly better
limits on $G_{E/M}^s$ will be needed in order to provide conclusive
evidence for or against the presence of a tangible non-perturbative
strange component in the nucleon.

%%%%%%%%%%%%%%%%%%%%%%%%%%%%%%%%%%%%%%%%%%%%%%%%%%%%%%%%%%%%%%%%%%%%%%%%%
\section{Future}

One can anticipate progress to be made on each of the issues addressed
here in the near future, as better quality data from high energy, high
luminosity facilities, capable of accessing extreme kinematic regions,
become available.
The semi-inclusive production of $\pi^\pm$ is an example of a 
straightforward way to cleanly extract the $d/u$ ratio at large $x$,
free of the nuclear contamination inherent in earlier analyses.
Once this is achieved, the origin of the SU(6) symmetry breaking
responsible for the softening of the $d$ quark distribution may be 
within reach.

% Another intriguing possibility to determine the free neutron
% structure function would be to use tritium targets to measure
% the $F_2^{^3H}/F_2^{^3He}$ structure function ratio.
% The nuclear effects in $^3H$ and $^3He$ would largely cancel
% in this combination, enabling a model-independence extraction
% of $F_2^n$ to within a few 1-2\%.

For the $\bar d/\bar u$ ratio, it is important experimentally to confirm
the downward trend of the ratio at large $x$, where pion cloud models
generally predict a flattening out rather than any dramatic decrease in
$\bar d/\bar u$.
Issues concerning the size of the gluon contribution at large $x$
may need to be resolved, however, before definitive conclusions
from the present data can be reached.
Future measurements of the neutron's electric form factor at
Jefferson Lab and elsewhere, as well as the $Q^2$-dependence
of the proton's form factors at large $Q^2$, should provide
critical tests of our understanding of the dynamics of light
quarks in the nucleon and the origin of the non-perturbative
spin-flavor interaction.

For the strange content of the nucleon, the data from CCFR continue
to be reanalyzed in view of possible nuclear shadowing corrections
and charm quark effects~\cite{BOROS}.
The strange form factors will also be measured to better precision
in the upcoming HAPPEX II experiment and subsequent experiments at
Jefferson Lab.

%%%%%%%%%%%%%%%%%%%%%%%%%%%%%%%%%%%%%%%%%%%%%%%%%%%%%%%%%%%%%%%%%%%%%%%%%
\section*{Acknowledgments}

This work was supported by the Australian Research Council.

%%%%%%%%%%%%%%%%%%%%%%%%%%%%%%%%%%%%%%%%%%%%%%%%%%%%%%%%%%%%%%%%%%%%%%%%%

\end{document}